\documentclass[aps,prb,showpacs,superscriptaddress,twocolumn,floatfix]{revtex4}

\usepackage{graphicx,bm}
\usepackage{amssymb,amsmath}
\usepackage{verbatim}
\usepackage{mathrsfs}

\begin{document}
\title{Spin projected unrestricted Hartree--Fock ground states for harmonic
  quantum dots.}

\date{\today}

\author{U. De Giovannini} 

\author{F.  Cavaliere}

\affiliation{Dipartimento di Fisica, Universit\`a di Genova, LAMIA
  CNR--INFM, Via Dodecaneso 33, 16146 Genova, Italy}

\author{R. Cenni} 

\affiliation{ Istituto Nazionale di Fisica Nucleare
  -- Sez.  Genova\\Dipartimento di Fisica, Universit\`a di Genova, Via
  Dodecaneso 33, 16146 Genova, Italy}

\author{M. Sassetti}

\affiliation{Dipartimento di Fisica, Universit\`a di Genova, LAMIA
  CNR--INFM, Via Dodecaneso 33, 16146 Genova, Italy}

\author{B.  Kramer}

\affiliation{I. Institut f\"ur Theoretische Physik, Universit\"at Hamburg, 
Jungiusstra\ss{}e 9 20355 Hamburg,\\ and Jacobs University Bremen, Campus Ring 1, 28759
  Bremen, Germany}

\begin{abstract}
  We report results for the ground state energies and wave functions
  obtained by projecting spatially unrestricted Hartree Fock states to
  eigenstates of the total spin and the angular momentum for harmonic
  quantum dots with $N\leq 12$ interacting electrons including a magnetic field. The ground
  states with the correct spatial and spin symmetries have lower
  energies than those obtained by the unrestricted method. The
  chemical potential as a function of a perpendicular magnetic field
  is obtained. Signature of an intrinsic spin blockade effect is found.
\end{abstract}

\maketitle
%%%%%%%%%%%%%%%%%%%%%%%%%%%%%%%%%%%%%%%%%%%%%%%%%%%%%%%%%%%%%%%%%%%%%%%%%%%
%
%
% SECTION: Introduction
%
%
%%%%%%%%%%%%%%%%%%%%%%%%%%%%%%%%%%%%%%%%%%%%%%%%%%%%%%%%%%%%%%%%%%%%%%%%%%%
\section{Introduction}
\label{introduction}
Systems like atoms, metal clusters, trapped bosons and quantum dots
show several universal features.~\cite{reimann} For example, strongly
interacting electrons in quantum dots arrange themselves in a rotating
Wigner molecule.~\cite{reimann} Rotating boson molecules have been
predicted to exist in ion traps.~\cite{RBM} Furthermore, symmetric
potentials can induce a shell structure in atoms,~\cite{bohrmottelson}
metal clusters,~\cite{clusterrmp} and quantum
dots.~\cite{kouwenhoven,reimann} In the latter, signatures of shell
structure have been experimentally probed,~\cite{tarucha,taruchasci}
leading to Hund's rules for the total spin of the electron ground
state. The spin in quantum dots~\cite{kouwenrev} also affects the
electron transport. It can lead to spin blockade
effects~\cite{weinmann,weinmann2} and negative differential
conductance in nonlinear
transport,~\cite{weinmann,cavaprlb,RHCSK2006,ciorgaapl,datta} and it
induces periodic modulations of the positions of the Coulomb peaks in
the linear conductance as a function of an applied magnetic
field.~\cite{roggesb,hawrylak3,hawrylak4,RHCSK2006} More recently, the
effect of the spatial distribution of the spins on the Kondo
phenomenon has been probed.~\cite{kondorog,roggekondo,kondokou,kondokel}

Electron and spin states of quantum dots have been theoretically
studied with various techniques.~\cite{reimann} For small electron
numbers $N$, exact diagonalization
(ED),~\cite{dineykhan,pfannkuche,merkt,mikhailov2,mikhailov,haw1,haw2,wojs,szafran,maksymB,kyriakidis,nishi}
configuration interaction (CI),~\cite{wensauer,rontani} and stochastic
variational methods~\cite{varga} allow for determining ground and
excited state energies and their quantum numbers with high accuracy.
For larger $N$, the size of the many-body basis set increases
exponentially. With presently available computational technology,
reliably converged ``exact'' results can be obtained only for electron
numbers up to $N\approx 8$ electrons.~\cite{rontani}

For $N\leq 13, N=16,24,48$, Quantum Monte Carlo
(QMC)~\cite{pederiva,egger,harju,bolton,ghosalnp,ghosal,pederiva2,esa}
methods have been used. They can provide accurate estimates for ground
and excited states energies. With these techniques, the shell
structure, Hund's rules, Wigner crystallization and the occurrence of
``magic'' angular momenta have been
investigated.~\cite{mikhailov,harju,maksym,ruan,filinov,bao,ghosalnp,ghosal}
Most of the results for higher particle numbers have been restricted
to zero magnetic field. It is believed that QMC provides better
estimates for the energies of the ground states for larger electron
numbers as compared to the ``exact'' methods.

For larger $N$ and/or in the presence of magnetic field $B$, 
methods like Hartree Fock (HF)
~\cite{landmanprl,landman3,reusch0,reusch,lipparini,hawrylak1,hawrylak2}
and the density functional
theory~\cite{manninen,hirose,gattobigio,harju2,esa1,esa2,esa3} have been used.
Generally, these seem to provide less accurate estimates for the
ground states which also can have unphysical broken symmetries due
to incomplete ansatz wave functions. For instance, neglecting
correlations, the straightforward HF method starts from a single Slater
determinant as a variational many-body wave function which not
necessarily is an eigenstate of the total spin.~\cite{szabo} Spatially
unrestricted HF methods~(UHF)\cite{landmanprl,reusch0} systematically
use symmetry breaking in order to obtain better estimates for the
ground state energy. This may lead to wrong results for the total
angular momentum and the total spin. For instance, UHF calculations
sometimes seem to fail predicting the total spin resulting from Hund's
rule, in contradiction to the more accurate methods. Violations of
Hund's rules for relatively weak Coulomb interactions have been
reported~\cite{landmanprl} for $N=4,8,9$.

Projection techniques, pioneered in the 60th of the last
century,~\cite{loewdin1,ring,loewdin2} can be applied for introducing
the correct spatial symmetries. In quantum dots, they have been used
for obtaining wave functions corresponding to specific angular
momenta.~\cite{landman2,landman4,landman5,koonin,YL2002} Recently, the
random phase approximation has been used to restore the rotational
symmetry of wave functions obtained by UHF.~\cite{serrarpa}

Restoring the {\em spin} symmetry has received much less attention,
and has been used only for very few (up to $N=3$)
electrons.~\cite{landman2} For $N=2$, the spin singlet symmetry has
been approximately restored with the Lipkin-Nogami
approach.~\cite{serrarpa} Larger $N$ have seldom been treated with the
projection technique.~\cite{degio} In view of the recent discussion of
spin effects in the transport spectra of quantum dots, information
about the total spin is, however, necessary.  Additionally, by
restoring the symmetries correlations are introduced into the ground
state wave function that are absent in a single UHF Slater
determinant. This leads to a better estimate for the ground state
energy.

In this paper, we apply a projection technique to the states obtained
by UHF for estimating the ground state energy of a circular quantum
dot with $N\leq 12$ electrons, including a magnetic field. Starting
from an UHF Slater determinant with broken rotational symmetry, a
first estimate for the ground state energy and the wavefunction is
obtained. Then, {\em both the total spin and the angular momentum} of
the UHF variational wave function are introduced by projecting on the
corresponding subspaces. We show that, after restoring {\em all of the
  symmetries}, the energies and the wave functions are improved and
show physical features which are not included in the UHF method.

We discuss the efficiency of the projected HF method (PHF) by
comparing our results with those of ED, CI, and QMC. We determine the
ground state energies as a function of a magnetic field, and obtain
the chemical potential that can be measured in transport experiments.
Our main findings are:

({\em i}) By projecting the UHF wave functions on the total angular
momentum $L$ and on the total spin $S$, the ground state energy is
successively lowered. The correction due to the spin projection is
generally smaller than the one associated with the angular momentum,
but still necessary for determining the correct ground state and its
quantum numbers.

({\em ii}) The quantum numbers $L$ and $S$ are correctly reproduced,
if the strength of the interaction is not too large. Especially, for
$B=0$, the first Hund's rule --- namely that $S$ is maximized for open
shells --- is recovered for $N\leq 12$ electrons, except for $N=10$,
 discussed below. Hund's rule has been
claimed earlier to be violated on the basis of UHF
results~\cite{landmanprl}.

({\em iii}) By comparing the results with CI and QMC, we estimate
a correlation energy, defined as the difference between PHF
and ``exact'' energies, of about 2\% of the ground state energy.

({\em iv}) With increasing interaction strength the correlation energy
decreases. Nevertheless, for stronger interaction, and larger $N$, the
PHF ground state tends to be spin polarized in contrast to more exact
results. This is consistent with earlier conjectures, namely that UHF
tends to overestimate the influence of the exchange.~\cite{reusch0}

({\em v}) As a function of $B$, several crossovers between ground
states with different total spins and angular momenta are found that
are absent in UHF. These are associated with characteristic changes in
the electron densities. The onset of the singlet--triplet
transition~\cite{hawrylak1} occurring for dot filling factor
$\nu\approx 2$ and $N$ even is recovered. Features that lead to an
intrinsic spin blockade are predicted.

In the next Section, details of the UHF method are outlined. The
consequences of the broken symmetries are described and the projection
technique is discussed, with special emphasis on the total electron
spin. In Sect.~\ref{results} results for zero and non-zero magnetic
field are presented and discussed. 
%%%%%%%%%%%%%%%%%%%%%%%%%%%%%%%%%%%%%%%%%%%%%%%%%%%%%%%%%%%%%%%%%%%%%%%%%%%%%%%
%
%
% SECTION: Model and Method
%
%
%%%%%%%%%%%%%%%%%%%%%%%%%%%%%%%%%%%%%%%%%%%%%%%%%%%%%%%%%%%%%%%%%%%%%%%%%%%%%%%
\section{Model and method}
\label{sectwo}
%%%%%%%%%%%%%%%%%%%%%%%%%%%%%%%%%%%%%%%%%%%%%%%%%%%%%%%%%%%%%%%%%%%%%%%%%%%%%%%
%
%
% SUBSECTION: The model
%
%
%%%%%%%%%%%%%%%%%%%%%%%%%%%%%%%%%%%%%%%%%%%%%%%%%%%%%%%%%%%%%%%%%%%%%%%%%%%%%%%
\subsection{The model}
\label{theham}
Consider $N$ electrons in a two-dimensional (2D) quantum dot confined
by an in--plane harmonic potential and subject to a perpendicular
magnetic field ${\bf B}=B{\bf e}_{z}$. The Hamiltonian is
($\hbar=c=1$)
\begin{equation}
  \label{hamil}
  {H}=\sum_{i=1}^{N}{h}_{0}({\bf r}_{i},s_{zi})+\frac{1}{2}\sum_{\substack{i,j=1\\i\neq j}}^{N}{v}({\bf r}_{i}-{\bf r}_{j})
\end{equation}
with
\begin{equation}
  \label{hamiltonian}
  {h}_{0}({\bf r},s_z)=\frac{{\left[{\bf p}+e{\bf A}({\bf
          r})\right]}^{2}}{2m^{*}}+\frac{m^{*}\omega_{0}^{2}}{2}{\bf
    r}^{2}+g^{*}\mu_{B}Bs_{z}
\end{equation}
${\bf r}\equiv(r,\vartheta)$ the 2D polar coordinates, $v({\bf
  r})={e^2}/{4\pi\varepsilon_{0}\varepsilon_{r} r}$ the Coulomb
interaction potential, ${\bf B}={\rm rot} {\bf A}$, $m^{*}$ effective
electron mass, $\omega_{0}$ confinement frequency, $g^{*}$ effective
$g$-factor and $\mu_{B}$ the Bohr magneton. The $z$-component of the
$i$-th spin is $s_{zi}=\pm 1/2$, $-e$ the electron charge and
$\varepsilon_{0}$ ($\varepsilon_{r}$) the vacuum (relative) dielectric
constant. The single-particle term in~(\ref{hamil}) yields the
Fock--Darwin~\cite{fd} (FD) spectrum
\begin{equation} 
\label{fdspect}
\epsilon_{n,l,s_{z}}=\Omega\left(2n+|l|+1\right)+
\frac{\omega_{c}}{2}l+g^{*}\mu_{B} B s_{z}\, ,
\end{equation}
with eigenfunctions $\phi_{n,l}({\bf r}) \chi^{\pm}$,
where $\chi^{+}$ ($\chi^{-}$) is the spinor corresponding to
$s_z=+1/2$ ($s_z= -1/2$) and~\cite{fd}
\begin{eqnarray}
  \phi_{n,l}({\bf r})&=& \frac{e^{il\vartheta}}{\ell_0}\sqrt{\frac{n!}{\pi
      (n+|l|)!}}\cdot\nonumber\\&&\cdot \left( \frac{r}{\ell_0}
  \right)^{|l|} e^{-\frac{r^{2}}{2\ell_0^2}}\mathscr{L}^{|l|}_{n}
\left( \frac{r^2}{\ell_0^2
    }\right)\, .\label{fdeigen} 
\end{eqnarray}
Here, $n$ and $l$ are principal and angular momentum quantum numbers,
$\ell_0=(m^*\Omega)^{-\frac{1}{2}}$ the characteristic oscillator
length and $\mathscr{L}_{n}^{|l|}(x)$ the generalized Laguerre
polynomial. The cyclotron frequency
$\omega_{c}=eB/m^{*}$, and the effective confinement frequency
$\Omega=(\omega_{0}^{2}+\omega_{c}^{2}/4)^{1/2}$ are introduced.

At $B=0$, expressing energies in units $\omega_{0}$ and lengths in
units $\ell_{0}$, the Hamiltonian~(\ref{hamil}) depends only on the
dimensionless parameter
\begin{equation}
\label{lambda}
\lambda=\frac{e^{2}}{4\pi\varepsilon_{0}\varepsilon_{r}\ell_{0}\omega_{0}}\,,
\end{equation}
which represents the relative strength of the interaction.
%%%%%%%%%%%%%%%%%%%%%%%%%%%%%%%%%%%%%%%%%%%%%%%%%%%%%%%%%%%%%%%%%%%%%%%%%%%%%%%
%
%
% SUBSECTION: The Unrestricted Hartree-Fock method
%
%
%%%%%%%%%%%%%%%%%%%%%%%%%%%%%%%%%%%%%%%%%%%%%%%%%%%%%%%%%%%%%%%%%%%%%%%%%%%%%%%
\subsection{The unrestricted Hartree-Fock method}
\label{HFteory}

In HF the Schr\"odinger equation for 
a given value of total $S_{z}=s_{z1}+\ldots+s_{zN}$ is solved by using orbitals
\begin{equation}
\label{orbitals}
\psi_{i}^{\alpha}({\bf r})=u_{i}^{\alpha}({\bf r})\chi^{\alpha}\, ,
\quad\quad\quad 1\leq i\leq N_{\alpha}
\end{equation}
with $\alpha=+$ ($\alpha=-$) denoting spin up (down) and 
$N_\alpha$ is the number of electrons with spin $\alpha\cdot 1/2$. They are
obtained as the solutions of the coupled integro--differential
equations
\begin{widetext}
\begin{equation}\label{HFE}
  \left[h_{0}({\bf r},\alpha 1/2)
+\int{\rm d}{\bf r}' \rho({\bf r}')v({\bf
      r}-{\bf r}')\right]
\psi_{i}^{\alpha}({\bf r})-
\sum_{j=1}^{N_{\alpha}}\left[\int{\rm d}{\bf r}'\psi_{j}^{\alpha *}
({\bf r}')\psi_{i}^{\alpha}({\bf r}') {v}({\bf r}-{\bf
      r}')\right]\psi_{j}^{\alpha}({\bf r})=\varepsilon_{i}^
{\alpha}\psi_{i}^{\alpha}({\bf r})\,,
\end{equation}
\end{widetext}
where $\rho({\bf r})=\rho^{+}({\bf r})+\rho^{-}({\bf r})$ is the HF density
\begin{equation}
\label{spresden}
  \rho^{\alpha}({\bf r})=\sum_{i=1}^{N_{\alpha}}
\left|\psi_{i}^{\alpha}({\bf r})\right|^2\, .
\end{equation}
For a given $S_{z}$, an initial guess for the orbitals
$\psi_{i}^{\alpha}({\bf r})$ with $i\leq N_{\alpha}$ is made. Then, HF
densities are evaluated and Eqs.~(\ref{HFE}) are solved to obtain
updated orbitals. This is iterated until self-consistency is achieved.
The many body wave function is a single Slater determinant,
eigenfunction of $S_{z}$,
\begin{eqnarray}\label{slater}
\left|\Psi^{S_{z}}\right>&=&\frac{1}{\sqrt{N!}}
{\rm det}\{\psi_{1}^{+}({\bf r}_{j}),\ldots,\psi_{N_{+}}^{+}({\bf r}_{j})\,,\nonumber\\
&&\qquad\qquad\psi_{1}^{-}({\bf r}_{j}),\ldots,\psi_{N_{-}}^{-}({\bf r}_{j})\}\, ,
\end{eqnarray}
that corresponds to a stationary point of the UHF
energy~\cite{szabo,ring}
\begin{equation}\label{HFfunctional}
E^{S_{z}}=\frac{\left<\Psi^{S_{z}}\right|H\left|\Psi^{S_{z}}\right>}
{\left<\Psi^{S_{z}}\right|\left.\Psi^{S_{z}}\right>}\,.
\end{equation} 

In order to numerically solve (\ref{HFE}) we expand the orbitals in
the FD basis $\phi_{\mu}({\bf r})\chi^{\alpha}$ (Eq. (\ref{fdeigen}))
\begin{equation}
\psi_i^{\alpha}({\bf r})=\sum_{\mu=1}^{K} C_{\mu i}^{\alpha}
\phi_\mu({\bf r})\chi^{\alpha}\, ,
\quad\quad\quad 1\leq \mu\leq K\label{exp1}
\end{equation}
where $C_{\mu i}^{\alpha}$ are complex coefficients. The truncation of
the basis to $K$ states is necessary in order to numerically implement
the procedure.  We have used the $K=75$ lowest FD states for each
value of $B$. This led to fair convergence (see
Sec.~\ref{somecomments}).  Introducing the density matrices
\begin{equation}
P_{\mu\nu}^{\alpha}=\sum_{i=1}^{N_{\alpha}}C_{\mu i}^{\alpha}(C_{\nu i}^{\alpha})^*\, ,\label{den1}
\end{equation}
connected to~(\ref{spresden}) by
\begin{equation*}
\rho^{\alpha}({\bf r})=\sum_{\mu,\nu=1}^{K}\phi_{\mu}({\bf r})P_{\mu\nu}^{\alpha}\phi_{\nu}^{*}({\bf r})\, ,
\end{equation*}
it is possible to show that equation~(\ref{HFE}) is equivalent to the coupled nonlinear
Pople--Nesbet eigenvalue problem
\begin{equation}
\sum_{\nu=1}^{K}F_{\mu \nu}^{\alpha}C_{\nu i}^{\alpha}= 
\varepsilon_{i}^{\alpha}C_{\mu i}^{\alpha}\, .\label{fock1}
\end{equation}    
Here $F_{\mu \nu}^{\alpha}$ are the Fock matrices,
\begin{eqnarray}
F_{\mu\nu}^{\alpha}&=&\epsilon_{n_{\mu},l_{\mu},\alpha{1}/{2}}\ 
\delta_{\mu\nu}\\
&&\!\!\!\!\!\!\!\!\!\!\!\!\!\!\!\!\!\!\!\! +\sum_{\lambda,\eta=1}^{K}
  P_{\lambda \eta}^{\alpha}\left[
    \langle\mu\eta|v|\nu\lambda\rangle-\langle\mu\eta|v|
\lambda\nu\rangle\right]+P_{\lambda\eta}^{-\alpha}
  \langle\mu\eta|v|\nu\lambda\rangle\, ,\nonumber
\end{eqnarray}
and the two-body interaction matrix elements
\begin{eqnarray}
  &&\!\!\!\!\!\!\langle\mu\eta|v|\nu\lambda\rangle=\nonumber\\&&\!\!\!\!\!\!
\int{\rm d}{\bf r}_1{\rm d}{\bf r}_2\phi_{\mu}^{*}({\bf r}_1)\phi_{\eta}^{*}
({\bf r}_2)v({\bf r}_{1}-{\bf r}_{2})\phi_{\nu}({\bf r}_1)\phi_{\lambda}({\bf r}_2)
\label{intermatel}
\end{eqnarray}
can be evaluated analytically.~\cite{rontani} The energy (\ref{HFfunctional}) is then
\begin{equation}
\label{EUHF}
  E^{S_{z}}=\frac{1}{2}\sum_{\alpha=\pm}\sum_{\mu,\nu=1}^{K}
\left[\epsilon_{n_{\nu},l_{\nu},\alpha{1}/{2}}\ 
\delta_{\nu\mu}+F_{\nu\mu}^{\alpha}\right]P_{\mu\nu}^{\alpha}\,.
\end{equation} 
We use {\em spatially unrestricted} initial
conditions~\cite{landmanprl,landman2,reusch0} with a random distribution
of initial $C^{\alpha}_{\mu,\nu}$. This implies
initial orbitals without circular symmetry,
and leads to better energy estimates. However, symmetry
broken Slater determinants are in general neither eigenfunctions of
the total angular momentum $L={l}_{1}+\ldots+{l}_{N}$ nor of ${\bf
  S}^{2}$ (total spin ${\bf S}={\bf s}_{1}+\ldots+{\bf
  s}_{N}$).~\cite{szabo}

The most general UHF solution is a linear superposition of
eigenfunctions $|\Psi^{S_{z}}(L,S)\rangle$ of $L$ and ${\bf S}$
\begin{equation}
\label{superposition}
|\Psi^{S_z}\rangle=\sum_{L=-\infty}^{\infty}\sum_{S\geq|S_z|}^{N/2}
|\Psi^{S_z}(L,S)\rangle\, .
\end{equation}

For given $N$ and $S_{z}$, many initial conditions are used.
Correspondingly, several stationary points are found. They form a
sequence $|\Psi_{k}^{S_{z}}\rangle$ ($k=1,2,\ldots$) with energies
$E_{1}^{S_{z}}<E_{2}^{S_{z}}<\ldots$.  For a given $S_{z}$, the
process is iterated until the lowest $E_{1}^{S_{z}}$ is found. The UHF
ground state is defined as,
\begin{equation*}
  E_{{\rm UHF}}=\mathop{{\rm min}}_{S_{z}}\ \left\{E_{1}^{S_{z}}\right\}\, .
\end{equation*}

%%%%%%%%%%%%%%%%%%%%%%%%%%%%%%%%%%%%%%%%%%%%%%%%%%%%%%%%%%%%%%%%%%%%%%%%%%%%%%%
%
%
% SUBSECTION: Spin and angular momentum projection
%
%
%%%%%%%%%%%%%%%%%%%%%%%%%%%%%%%%%%%%%%%%%%%%%%%%%%%%%%%%%%%%%%%%%%%%%%%%%%%%%%%
\subsection{Spin and angular momentum projection}
\label{project}
In order to obtain states with specific $L$ and $S$ we act on the UHF
Slater determinant with operators~\cite{ring} $\hat{P}_{L}$ and
$\hat{P}_{S}^{S_z}$ which project on $\hat{L}$ and $\hat{{\bf S}}$,
respectively. They satisfy commutation rules
$[\hat{P}_{S}^{S_z},\hat{P}_{L}]=[\hat{P}_{S}^{S_z},\hat{H}]=[\hat{P}_{L},\hat{H}]=0$.
Their simultaneous action yields an eigenfunction of $\hat{L}$ and
$\hat{{\bf S}}^{2}$,
$\hat{P}_{L}\hat{P}^{S_z}_{S}|\Psi^{S_z}\rangle=|\Psi^{S_z}(L,S)\rangle$.
The corresponding energy is  (Appendix~\ref{details})
\begin{eqnarray}
  E^{S_{z}}(L,S)&=&\frac{\langle\Psi^{S_z}(L,S)|\hat{H}|\Psi^{S_{z}}(L,S)\rangle}
  {\langle\Psi^{S_z}(L,S)|\Psi^{S_{z}}(L,S)\rangle}\nonumber\\
  &\equiv&
  \frac{\langle\Psi^{S_z}|\hat{H}|\Psi^{S_{z}}(L,S)\rangle}
  {\langle\Psi^{S_z}|\Psi^{S_{z}}(L,S)\rangle}\label{energyp}\, . 
\end{eqnarray} 

The spin projector
\begin{equation}
\label{projs}
\hat{P}_{S}^{S_z}=\prod_{k=|S_{z}|,k\neq S}^{N/2}\frac{\hat{S}^{2}-k(k+1)}{S(S+1)-k(k+1)}\ ,
\end{equation}
 annihilates all the components of~(\ref{slater}) with spin
different from $S$.~\cite{loewdin1}  Its action is written as~\cite{loewdin1,projS1}
\begin{equation}
\label{project1}
\hat{P}_{S}^{S_z}|\Psi^{S_z}\rangle=\sum_{q=0}^{N_{<}}C_{q}(S,S_{z},N)|T_{q}\rangle
\end{equation}
where $N_{<}={\rm min}\{N_{+},N_{-}\}$ and 
\begin{widetext}
\begin{equation}\label{sanibel}
C_{q}(S,S_{z},N)=\frac{2S+1}{1+N/2+S}\sum_{k=0}^{S-S_z}(-1)^{q+S-S_z-k}
{S-S_z\choose k}{S+S_z\choose S-S_z-k}
{N/2+S\choose S_{z}+N/2-q+k}^{-1} 
\end{equation}
\end{widetext}
are the Sanibel coefficients.~\cite{projS1,ruitz} The term
$|T_{q}\rangle=|T_{q}^{(1)}\rangle+\ldots+|T_{q}^{(n_q)}\rangle$ is the
sum of all
\begin{equation}
\label{numberofterms}
n_q={N_{+} \choose q}{N_{-} \choose q}
\end{equation}
Slater determinants obtained by interchanging, without
repetition, {\em all} the possible $q$ spinor pairs with opposite
spins in $|\Psi ^{S_z}\rangle$. By definition
$\left|T_0\right>\equiv |\Psi^{S_z}\rangle$.

For example, consider $N=4$, $S_z=0$
($N_+=N_-=2$),
\begin{equation}
\left| T_0 \right>=\frac{1}{\sqrt{24}}
{\rm det}\{u_{1}^{+}{\chi^+},u_{2}^{+}{\chi^+},u_{1}^{-}{\chi^-},u_{2}^{-}{\chi^-}\}\, .
\end{equation} 
This state is a linear superposition of all spin eigenstates with
$S\leq 2$. The spin projection selects a specific spin
\begin{eqnarray}\label{san1}
&&\hat{P}_{S=0}^{S_z=0}\left|\Psi^{S_{z}=0}\right>=\frac{1}{3} 
|T_0\rangle-\frac{1}{6} |T_1\rangle +\frac{1}{3}|T_2\rangle\, ,\\ 
\label{san2}
&&\hat{P}_{S=1}^{S_z=0}\left|\Psi^{S_{z}=0}\right>= \frac{1}{2} 
|T_0\rangle - \frac{1}{2}|T_2\rangle\, ,\\
\label{san3}
&&\hat{P}_{S=2}^{S_z=0}\left|\Psi^{S_{z}=0}\right>= \frac{1}{6} 
|T_0\rangle +\frac{1}{6} |T_1\rangle + \frac{1}{6}|T_2\rangle\, ,
\end{eqnarray}
where
\begin{eqnarray*}
|T_{1}\rangle&=&\frac{1}{\sqrt{24}}\left[{\rm
  det}\{u_{1}^{+}{\chi^+},u_{2}^{+}{\chi^-},u_{1}^{-}{\chi^+},u_{2}^{-}{\chi^-}\}\right.+\\
&+&{\rm
  det}\{u_{1}^{+}{\chi^+},u_{2}^{+}{\chi^-},u_{1}^{-}{\chi^-},u_{2}^{-}{\chi^+}\}+\\
&+&{\rm
  det}\{u_{1}^{+}{\chi^-},u_{2}^{+}{\chi^+},u_{1}^{-}{\chi^+},u_{2}^{-}{\chi^-}\}+\\
&+&\left.{\rm
  det}\{u_{1}^{+}{\chi^-},u_{2}^{+}{\chi^+},u_{1}^{-}{\chi^-},u_{2}^{-}{\chi^+}\}\right]\, ,\\
|T_{2}\rangle&=&\frac{1}{\sqrt{24}}{\rm
  det}\{u_{1}^{+}{\chi^-},u_{2}^{+}{\chi^-},u_{1}^{-}{\chi^+},u_{2}^{-}{\chi^+}\}\, .\\
\end{eqnarray*}
Summing up equations (\ref{san1})---(\ref{san3}) results in the
original determinant $|\Psi ^{S_z}\rangle$, since $\sum_{S}
\hat{P}_S^{S_z}=1$.

The projector on $\hat{L}$ is given by~\cite{ring}
\begin{equation}
\label{projl}
\hat{P}_{L}=\frac{1}{2\pi}\int_{0}^{2\pi}{\rm d}\gamma\ e^{-iL\gamma}e^{i\hat{L}\gamma}\, ,
\end{equation}
where $\exp{(i\hat{L}\gamma)}$ acts on $|T_{q}\rangle$ rotating by
$\gamma$ around the $z$ axis all spatial parts of the orbitals
\begin{equation*}
u_{i}^{\alpha}(r,\vartheta)\rightarrow u_{i}^{\alpha}(r,\vartheta+\gamma)\, .
\end{equation*}
We denote this by $|T_{q}(\gamma)\rangle$.
Using~(\ref{project1}) and~(\ref{projl}) we get
\begin{eqnarray}
\nonumber
|\Psi^{S_{z}}(L,S)\rangle&=&\frac{1}{2\pi}
\sum_{q=0}^{N_{<}}C_{q}(S,S_{z},N)\cdot\\&&\cdot\int_{0}^{2\pi}{\rm
  d}\gamma\ e^{-iL\gamma}|T_{q}(\gamma)\rangle\, .\label{project2}
\end{eqnarray} 
The projected state~(\ref{project2}) is a sum of many Slater
determinants  (Appendix~\ref{details}). This indicates that correlation has been
introduced by the projection.

The main computational effort is due to the evaluation of two-body
matrix elements in~(\ref{energyp}). Projecting an $N$-particle UHF
state with $S_z$ to a state with total spin $S$ requires to evaluate
$n({S_z},N)=\sum_{q=0}^{N_<}n_{q}$ terms,
\begin{equation}\label{numot}
  n(S_{z},N)={N \choose N/2+S_{z}}.
\end{equation}
For $N$ even (odd), the worst case is $S_z=0$ ($S_z=1/2$).

For the angular momentum projection we use a fast Fourier transform
(FFT) and partition the integration interval $[0,2\pi]$ in $n(L)$
points; $n(L)$ is determined by the angular
momentum range $|L|\leq L_{max}$ for which good convergence (relative
error $< 10^{-6}$) of the PHF energies is required. We have checked
that for $L_{max}=20$ $n(L)=256$ is needed. Using FFT,
all energy values for given $S$ and $|L|\leq L_{max}$ are {\em
  simultaneously} available, which considerably accelerates the
calculation with respect to performing distinct computations
for each value of $L$. The total number of two-body matrix elements
is $n_{tot}=n({S_z},N)n(L)$.
\begin{table}[htbp]
\begin{ruledtabular}
\begin{tabular}{rrr}
{$N$}&{$n_{tot}$}&$n$\\
\hline\hline
{2}&512&\\
{4}&1536&19774\cite{szafran}\\
{6}&5120&661300\cite{rontani}\\
{8}&17920&\\
{10}&64512&\\
{12}&236544&\\
{14}&878592&\\
{16}&3294720&
\end{tabular}
\end{ruledtabular}
\caption{Numbers of two-body matrix elements used for PHF, $n_{tot}$ 
required to evaluate $E^{S_{z}=0}(L,S)$ for even 
$N\leq 16$, $S_{z}=0$, $S\leq N/2$, and $|L|\leq 20$ 
(see text). Last column:
some numbers used in other methods. \label{tab0}}
\end{table}

Table~\ref{tab0} shows $n_{tot}$ for the case of even $N\leq 16$ and
$L_{max}=20$ in the worst case $S_z=0$. This is the $L_{max}$ value used in the paper.
Although $n_{tot}$ quickly increases as a function of $N$, especially
because~(\ref{numot}) grows exponentially for large $N$, it still
compares favorably with respect to exact methods. For example,
previously reported ED calculations~\cite{szafran} used a basis of
19774 Slater determinants for $N=4$, $S_z=0$, $S=2$, and
$L=14$. CI calculations~\cite{rontani} for $N=6$, $S_z=0$,
$S=0$, $L=0$ need 661300 configurational state functions (linear
superposition of Slater determinants). 
%%%%%%%%%%%%%%%%%%%%%%%%%%%%%%%%%%%%%%%%%%%%%%%%%%%%%%%%%%%%%%%%%%%%%%%%%%%%%%%
%
%
% SUBSECTION: Determining the PHF ground state
%
%
%%%%%%%%%%%%%%%%%%%%%%%%%%%%%%%%%%%%%%%%%%%%%%%%%%%%%%%%%%%%%%%%%%%%%%%%%%%%%%%
\subsection{Determining the PHF ground state}
To determine the ground state, it is generally {\em not} sufficient to
project only the UHF ground state. If several UHF solutions
(Sec.~\ref{HFteory}) are almost degenerate, all of the
$|\Psi_{i}^{S_{z}}\rangle$ have to be projected. The PHF ground state
is then defined by
\begin{equation}\label{ephf}
  E_{{\rm PHF}}=\mathop{{\rm min}}_{\{i,S_{z},L,S\}}\ 
\left\{E_{i}^{S_{z}}(L,S)\right\}\, .
\end{equation}
One can show that projecting arbitrary UHF Slater determinants on $L$
and $S$ always leads to energies that are {\em not lower} than the
exact ground state energy, thus satisfying the variational principle.

As an example, we consider $N=4$ for $B=0$, with confinement energy
$\omega_{0}=0.741$ meV. We assume the standard GaAs parameters,
$m^{*}=0.067\ m_e$ and $\varepsilon_r=12.4$. The confinement
corresponds to $\lambda=4$. For each $S_{z}$, we have used more than
500 UHF initial conditions. We found two solutions with $S_{z}=0$,
and one for $S_{z}=1$ and $S_{z}=2$. The
corresponding energies $E^{S_z}_{i}$ are given in Tab.~\ref{tab1}.
\begin{table}[htbp]
\begin{ruledtabular}
\begin{tabular}{cl|lc}
{$S_z$} &{$E_{i}^{S_{z}}$}&{$\!\!E_{i}^{S_{z}}(L,S)$}&$L$ $S$\\
\hline\hline
{ 0}&19.612    &    19.356 &0 0\\
{}  &{}        &    19.404 &1 1 \\
\cline{2-4}
{}  &19.641    &    $19.331^\star$ &0 1\\       
{}  &{}        &    19.515 &2 0\\
\hline
{ 1}&19.608    &    19.342 &0 1\\
{}  &          &    19.394 &1 1\\
\hline
{ 2}&19.581$^{\star}$    &    19.516 &2 2\\
\end{tabular}
\end{ruledtabular}
\caption{\label{tab1} PHF  for $N=4$ for a GaAs 
quantum dot with $B=0$, $\omega_{0}=0.741$ meV, 
 $m^*=0.067\ m_e$,
 $\varepsilon_r=12.4$,  $\lambda=4$; 
2nd column: UHF energy $E_{i}^{S_z}$ (units $\omega_0$); 
3rd column: PHF energy $E_{i}^{S_z}(L,S)$ (units $\omega_0$); 4th column:
PHF quantum numbers. Ground states are indicated by $^\star$.}
\end{table}

The UHF ground state corresponds to $S_{z}=2$. Applying the projection
to each UHF state, we obtain $E_i^{S_z}(L,S)$ with different quantum
numbers $L$ and $S$: the lowest two are given. The PHF ground state
corresponds to $L=0$ and $S=1$. The latter is obtained by projecting
the energetically higher UHF minimum with $S_z=0$.
%%%%%%%%%%%%%%%%%%%%%%%%%%%%%%%%%%%%%%%%%%%%%%%%%%%%%%%%%%%%%%%%%%%%%%%%%%%%%%%
%
%
% SUBSECTION: Some comments about errors
%
%
%%%%%%%%%%%%%%%%%%%%%%%%%%%%%%%%%%%%%%%%%%%%%%%%%%%%%%%%%%%%%%%%%%%%%%%%%%%%%%%
\subsection{Some comments about errors}
\label{somecomments}
The major systematic error of the UHF approximation is the
neglect of correlations.
By projecting the slater determinant on fixed angular momentum and spin  PHF
attempts to correct for these effects. A second systematic effect is
due to the uncertainty if the self consistent HF procedure has
converged towards the absolute minimum of the energy.

In determining the UHF ground state energies, we have checked that the
convergence with respect to the size of the basis set is better than
$10^{-6}$.

For getting insight into the above systematic effects one can start from
wave functions with the same $L,S$ but originating from UHF states with
different $S_z$. They should be degenerated at $B=0$. In the example
of Tab.~\ref{tab1}, these are the pairs $|\Psi_{1}^{0}(1,1)\rangle$,
$|\Psi_{1}^{1}(1,1)\rangle$ and $|\Psi_{2}^{0}(0,1)\rangle$,
$|\Psi_{1}^{1}(0,1)\rangle$. Their energetic differences are $0.010\ 
\omega_0$ and $0.011\ \omega_0$, respectively. This corresponds to a
relative uncertainty of $5\cdot 10^{-4}$. Similar estimates for the
``degeneracy error'' is obtained from data for different $N$ and
$\lambda$. We attribute the degeneracy error mainly to UHF: different
UHF states in different $S_{z}$ sectors approximate the true states
with different precision. Therefore, their projection on the same
$L,S$ sector does not yield exactly degenerate states.

By comparing our results with other works (see below), the PHF ground
state energies for $N\geq 3$ remain about 2\% higher than those
obtained with ED and QMC. This can be attributed to correlations
beyond those introduced by the projection. This also is the limiting
factor for the ground state quantum numbers in the regime $N\geq 6$
and $\lambda\geq 4$ where too high polarization are obtained.

When several PHF energies $E^{S_z}_{i}(L,S)$ are almost degenerate,
one can improve further the ground state: linear superposition of the
almost degenerate states $|\Psi^{S_z}_{i}(L,S)\rangle$ may result in
further lowering of the energy. Here, we have not systematically
investigated this effect.
%%%%%%%%%%%%%%%%%%%%%%%%%%%%%%%%%%%%%%%%%%%%%%%%%%%%%%%%%%%%%%%%%%%%%%%%%%%%%%%
%
%
% SECTION: Results
%
%
%%%%%%%%%%%%%%%%%%%%%%%%%%%%%%%%%%%%%%%%%%%%%%%%%%%%%%%%%%%%%%%%%%%%%%%%%%%%%%%
\section{Results}
\label{results}
%%%%%%%%%%%%%%%%%%%%%%%%%%%%%%%%%%%%%%%%%%%%%%%%%%%%%%%%%%%%%%%%%%%%%%%%%%%%%%%
%
%
% SUBSECTION: Zero magnetic field
%
%
%%%%%%%%%%%%%%%%%%%%%%%%%%%%%%%%%%%%%%%%%%%%%%%%%%%%%%%%%%%%%%%%%%%%%%%%%%%%%%%
\subsection{Zero magnetic field}
\label{bzero}
%%%%%%%%%%%%%%%%%%%%%%%%%%%%%%%%%%%%%%%%%%%%%%%%%%%%%%%%%%%%%%%%%%%%%%%%%%%%%%%
%
%
% SUBSUBSECTION: Ground state energies
%
%
%%%%%%%%%%%%%%%%%%%%%%%%%%%%%%%%%%%%%%%%%%%%%%%%%%%%%%%%%%%%%%%%%%%%%%%%%%%%%%%
\subsubsection{Ground state energies}
\label{energies}
\begin{table}[htbp]
\begin{ruledtabular}
\begin{tabular}{cc|c|c|c|c|c}
$N$ & $\lambda$ & $E_{\rm PHF}$ & $L$ $S$& $E_{\rm CI}$ &$E_{\rm DMC}$  &$L$ $S$\\
\hline\hline
2   & 1.89      & 3.817         & 0 0    &         &3.649&0 0    \\
    & 2         & 3.885         & 0 0    & 3.7295  &     &0 0    \\
    & 4         & 4.983         & 0 0    & 4.8502  &    &0 0    \\
\hline
3   & 1.89      & 8.154         & 1 1/2  &         &7.978& 1 1/2   \\
    &2          & 8.337         & 1 1/2  & 8.1671  &     & 1 1/2  \\
    & 4         & 11.131       & 0 3/2  &11.043    &     & 1 1/2 \\
\hline           
4   & 1.89      & 13.554        & 0 1    &         &13.266& 0 1  \\
  & 2         & 13.899        & 0 1    &13.626     &     &  0 1 \\
  & 4         & 19.330        & 0 1    &19.035     &     &  0 1  \\
\hline 
5 & 1.89      & 20.264        & 1 1/2  &           &19.764& 1 1/2  \\
  & 2         & 20.811        & 1 1/2  &20.33      &       & 1 1/2 \\
  & 4         & 29.501        & 1 1/2  &28.94      &       & 1 1/2  \\
\hline          
6 & 1.89      &27.905         & 0 0    &           &27.143& 0 0  \\
  & 2          & 28.703        & 0 0   &27.98     &       & 0 0  \\
  & 4          & 41.187       & 0 3   &40.45     &       & 0 0   \\
\hline
7 & 1.89       & 36.627        & 2 1/2 &          &35.836 &2 1/2  \\ 
  & 2          & 37.698        & 2 1/2 &          &       &      \\
  & 4          & 54.497      & 0 5/2  &(54.68)     &53.726   & (2)2 (1/2)1/2\\
\hline
8 & 1.89        & 46.260       & 0 1    &          & 45.321& 0 1   \\
   & 2          & 47.659       & 0 1    & 47.14  & 46.679  & 0 1  \\
   & 4          & 69.479       & 0 4    & 70.48   &        & 0 1 \\
\hline 
9  & 1.89       & 56.853       & 0 3/2  &          & 55.643& 0 3/2  \\
\hline          
10 & 1.89    & 68.245          & 0 0         &          &   66.8785 & 2 1   \\
   &         & (68.283)         & 2 1         &          &   (66.8789) & 0 0   \\
\hline
11 & 1.89    & 80.444        & 0 1/2      &            &  78.835   & 0 1/2  \\ 
\hline
12 & 1.89     & 93.661        & 0 0       &             &  91.556  & 0 0   \\
\end{tabular}
\end{ruledtabular}
\caption{\label{tab2} Ground state energies from PHF 
for $N\leq 12$ and $\lambda=1.89$, $2$, $4$
with corresponding $L$, $S$ 
($m^*=0.067\ m_e$, and $\varepsilon_r=12.4$)
together with results from
 CI~\cite{rontani}, 
and DMC (Ref.~\onlinecite{pederiva} for $\lambda
  =1.89$, Ref.~\onlinecite{ghosal} for $\lambda\geq 2$). All energies are in units $\omega_0$.
 }
\end{table}

Table~\ref{tab2} summarizes our results for the ground state energies
at $B=0$, for $N\leq 12$ and $\lambda=1.89$, $2$, $4$.  Results
obtained with Diffusion Monte Carlo (DMC)~\cite{pederiva,ghosal} and
CI~\cite{rontani} are included.

For $\lambda\leq2$, angular momenta and total spins of the ground states obtained by PHF
agree with DMC and CI. The total spin fulfills Hund's first rule: a
singlet state for the filled shells ($N=2,6,12$), a triplet for
$N=4,8$ and $S=3/2$ for $N=9$. Only for $N=10$, Hund's rule is not
fulfilled since we find $S=0$ instead of $S=1$. However, here the
degeneracy error is $0.064\, \omega_0$, larger than the energy distance
$\Delta E=0.038\, \omega_0$ between the ground and the first excited
state.  Also DMC~\cite{pederiva} predicts an extremely small energy
gap between the singlet and the triplet, though it yields an $S=1$
ground state.

Increasing the interaction strength ($\lambda=4$) PHF still produces
energies consistent with CI and DMC. However, for $N=3,6,7,8$
incorrect quantum numbers are predicted with a tendency towards
polarization. Whenever polarization occurs the ground states have low
angular momenta in PHF.

For $\lambda > 4$, preliminary results indicate deviations of PHF with
respect to CI, DMC. They are reminiscent of the tendency
of HF to predict spin polarized ground states due to overestimating
the exchange as compared to correlations.

\begin{figure}[htbp]
\includegraphics[width=8cm,keepaspectratio]{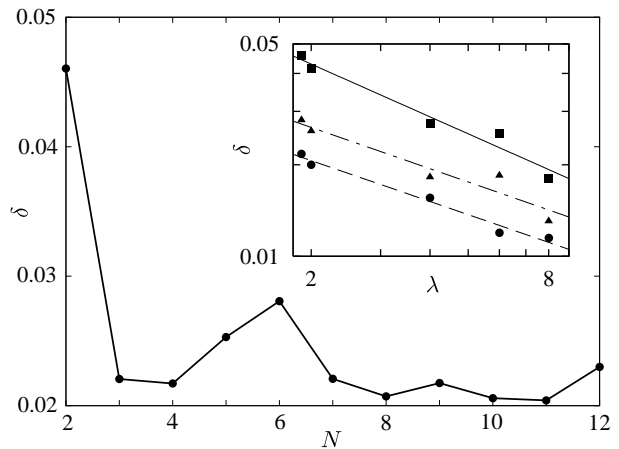}
\caption{\label{fig:error} Deviations
  $\delta$ between PHF and DMC~\cite{pederiva} for $2\leq N\leq 12$,
  with $\lambda=1.89$ (Tab.~\ref{tab2}). Inset: double logarithmic
  plot of $\delta({\lambda})$ from PHF and QMC\cite{pederiva}
  ($\lambda=1.89$), QMC~\cite{ghosal} and CI~\cite{rontani}
  ($\lambda\geq 2$) for $N=2$ (squares), $N=4$ (dots) $N=6$ (triangles).
  Lines: best fits to data.}
\end{figure}

The relative deviation $\delta=(E_{\rm PHF}-E_{\rm DMC})/E_{\rm DMC}$
for\cite{pederiva} $\lambda=1.89$ and $2\leq N\leq 12$ is shown in
Fig.~\ref{fig:error}; $\delta$ is largest for the closed shells
$N=2,6,12$. Except for $N=2$, $\delta\approx 2$\%. The inset shows
$\delta$ for $N=2$ (squares, with $L,S=0,0$), $N=4$ (dots, with $L,S=0,1$), 
and $N=6$ (triangles, with $L,S=0,0$) within $1.89\leq\lambda\leq8$. A decrease with
$\lambda$ according to a power law is observed, $\lambda^{-\beta(N)}$.
By numerically fitting the data, one finds
$\beta(2)=0.57$, $\beta(4)=0.44$ and $\beta(6)=0.45$.

\begin{table}[htbp]
\begin{ruledtabular}
\begin{tabular}{cc|cc|cc|c}
$E_{i}^{S_{z}}$   & $S_z$ & $E_{i}^{S_z}(L)$ & $L$ 
& $E_{i}^{S_z}(L,S)$ &  $S$& $E_{\rm DMC}$  \\
\hline\hline
48.150          & 0      &47.842           & 0   &47.659$^0$     & 1     &46.679 \\
                &        &                 & 0   &48.031$^1$     & 2     & \\
                &        &48.088           & 2   &47.790$^{\ }$    & 0     &46.875 \\
                &        &                 & 2   &47.799$^4$     & 1     & \\
                &        &47.971           & 1   &47.817$^2$     & 2     &46.917 \\
                &        &                 & 1   &48.028$^3$     & 1     &       \\
                &        &48.076           & 4   &47.777$^{\ }$    & 0     &46.779 \\
48.237          & 0      &47.981           & 0   &47.805$^{\ }$    & 0     &46.807 \\
                &        &48.025           & 1   &47.910$^3$     & 1     &       \\
\hline
48.131          & 1      &47.796           & 2   &47.742$^4$     & 1     &46.756 \\
                &        &47.887           & 1   &47.806$^2$     & 2     &46.917 \\
                &        &                 & 1   &47.985$^3$     & 1     &       \\
                &        &48.022           & 0   &47.977$^1$     & 2     &47.406 \\
                &        &                 & 0   &47.997$^0$     & 1     &       \\
\hline
48.243          & 2      &47.896           & 1   &47.881$^2$     & 2     & \\
\hline
48.335          & 3      &48.129           & 3   &48.126$^{\ }$    & 3     &47.404 \\
\end{tabular}
\end{ruledtabular}
\caption{\label{tab3} Comparison of the lowest energies $E_{i}^{S_{z}}$
  obtained from UHF, 
followed by projection on angular momentum, $E_{i}^{S_z}(L)$, 
and total spin, $E_{i}^{S_z}(L,S)$ 
for $N=8$, $\lambda=2$ 
($m^*=0.067\ m_e$, $\varepsilon_r=12.4$). 
Last column: energies from
  DMC.~\cite{ghosal} The ground state has $L,S=0,1$. 
Superscripts $^{0,1,2,3,4}$ denote ``degenerated'' energies
 with the same quantum numbers $L,S$ but originating from different
  $S_z$. All energies are in units $\omega_0$.}
\end{table}

\begin{figure}[htbp]
\includegraphics[width=8cm,keepaspectratio]{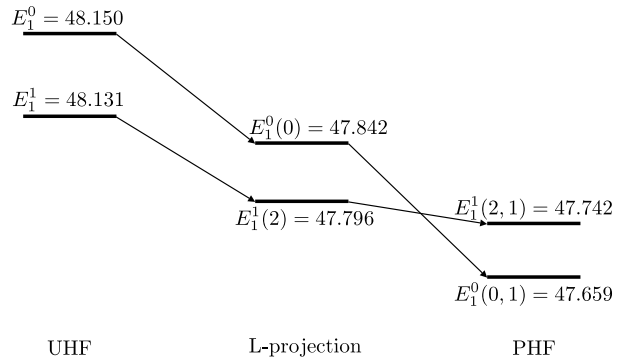}
\caption{\label{levels2} Influence of the projection procedure on the
  energy levels (unit $\omega_0$) for $N=8$ and $\lambda=2$.  Only the
  two lowest UHF states (left) directly involved in the determination
  of the PHF ground state (right) are shown.}
\end{figure}

Table~\ref{tab3} and Fig.~\ref{levels2} illustrate the effect of
angular momentum projection alone followed by spin projection for
$N=8$ starting with UHF states with $S_z=0,\ldots,3$. The energy
$E_{i}^{S_{z}}(L)$ projected on angular momentum
is
\begin{equation}
\label{energyL}
E_{i}^{S_{z}}(L)=\frac{\langle\Psi_{i}^{S_{z}}|H|
\Psi_{i}^{S_{z}}(L)\rangle}
{\langle\Psi_{i}^{S_{z}}|\Psi_{i}^{S_{z}}(L)\rangle}\,,
\end{equation}
where $|\Psi_{i}^{S_{z}}(L)\rangle=P_{L}|\Psi_{i}^{S_{z}}\rangle$.
Only the lowest energies are included in the table. 

The typical energy gain obtained by angular momentum projection is
about $0.25\,\omega_0$. The spin projection induces corrections of
the same order of magnitude, which can even change the sequence of
energies (Fig.~\ref{levels2}).
From the UHF state with $S_z=0$ and $E_{1}^{0}=48.150\, \omega _0$,
which is {\em not} the UHF ground state, projection on $L=0$ yields
$E_{1}^{0}(L=0)=47.842\, \omega _0$.  After projection on
the total spin we obtain the energy of the ground state,
$E_{1}^{0}(L,S=0,1)=47.659\, \omega _0$ and an excited state at
$E_{1}^{0}(L,S=0,2)=48.031\, \omega _0$. On the other hand, the
energetically lowest UHF minimum $E_{1}^{1}=48.131\, \omega_0$ turns
out to yield the first excited PHF state at
$E_{1}^{1}(L,S=2,1)=47.742\, \omega _0$.

Thus, PHF not only introduces a lowering of the energies but can also
restore the correct ordering of energy levels. This can be seen from
the last column of Tab.~\ref{tab3}, which contains the results
obtained by DMC~\cite{ghosal}. Restoring the spin plays a crucial role
in obtaining {\em all} correct quantum numbers for the ground state
including Hund's rule.~\cite{degio} For example with angular momentum projection
alone, one would have predicted $L=2$ for the ground state, in
contrast to the correct result.

The degeneracy error for this case is approximately $0.06\ \omega_0$
(some example of almost degenerate states are included in
Tab.~\ref{tab3}). The distance between ground state and the first
excited state is $\approx\ 0.08\ \omega_0$. This suggests that the
ground state for $N=8$ has $L,S=0,1$, consistent with DMC. Even the
quantum numbers of the first three excited states turn out to be
reproduced correctly while the 4th and the 5th appear to be
interchanged.

%%%%%%%%%%%%%%%%%%%%%%%%%%%%%%%%%%%%%%%%%%%%%%%%%%%%%%%%%%%%%%%%%%%%%%%%%%%%%%%
%
%
% SUBSUBSECTION: Ground state densities
%
%
%%%%%%%%%%%%%%%%%%%%%%%%%%%%%%%%%%%%%%%%%%%%%%%%%%%%%%%%%%%%%%%%%%%%%%%%%%%%%%%
\subsubsection{Ground state densities}
\label{densities}
\begin{figure}[htbp]
\includegraphics[width=8cm,keepaspectratio]{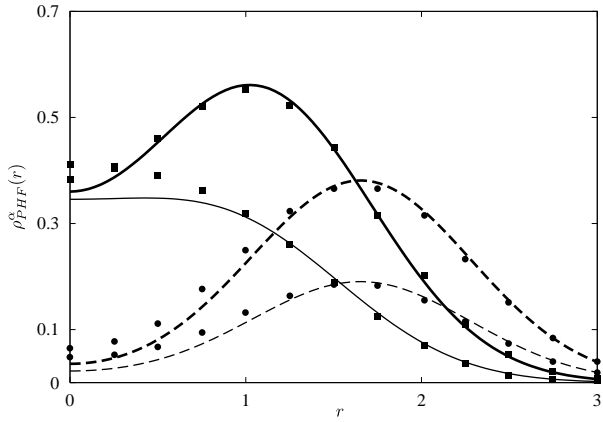}
\caption{\label{den3} Spin resolved densities $\rho_{{\rm
      PHF}}^{\alpha}(r)$ (thick line: $\alpha=+$; thin line:
  $\alpha=-$) for a GaAs quantum dot with $N=3$, $L=1$, $S=1/2$, and
  $S_{z}=1/2$ for interaction strengths $\lambda=2$ (solid),
  $\lambda=8$ (dashed). Density unit: $\pi^{-1}\ell_0^{-2}$. Data from
  ED:~\cite{mikhailov1} squares $\lambda=2$, circles $\lambda=8$.}
\end{figure}

\begin{figure}[htbp]
\includegraphics[width=8cm,keepaspectratio]{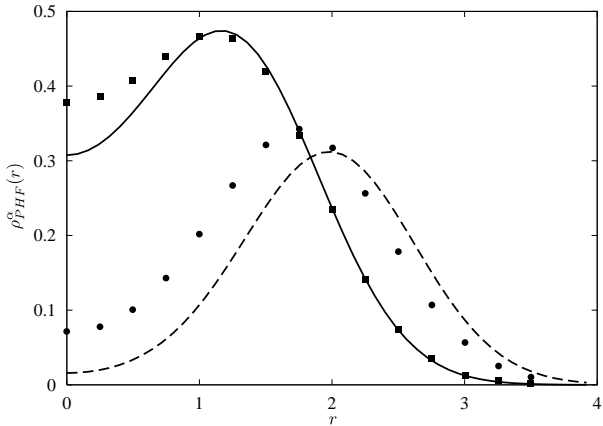}
\caption{\label{den4}Spin resolved densities $\rho_{{\rm
      PHF}}^{+}(r)=\rho_{{\rm PHF}}^{-}(r)$ for a GaAs quantum dot
  with $N=4$, $L=1$, $S=1$, and $S_z=0$ for $\lambda=2$ (solid line),
  $\lambda=8$ (dashed line).  Density unit: $\pi^{-1}\ell_0^{-2}$. Data from
  ED:~\cite{mikhailov} squares $\lambda=2$, circles $\lambda=8$.}
\end{figure}

For the spin-resolved densities 
\begin{equation*}
\rho_{{\rm PHF}}^{\alpha}(r)=
\frac{
\langle \Psi^{S_z}(L,S)|\sum_{i=1}^{N_{\alpha}}\delta({\bf r}-{\bf r}_{i})
\delta_{s_{zi},\alpha\cdot 1/2}|\Psi^{S_z}(L,S)\rangle}
{\langle \Psi^{S_z}(L,S)|\Psi^{S_z}(L,S)\rangle}\,
\end{equation*}
we first consider $N=3$ and $N=4$ (Figs.~\ref{den3} and~\ref{den4})
for intermediate ($\lambda=2$) and strong ($\lambda=8$) interaction.
Increasing the interaction strength leads to a shift of the maximum of
the densities towards higher $r$, consistent with earlier findings by
ED.~\cite{mikhailov,mikhailov1} This is clearly observed in the spin
up density for $N=3$ (Fig.~\ref{den3}). For $N=4$, the ground state
($L=0$, $S=1$, $S_z=0$) densities $\rho_{{\rm PHF}}^{+}(r)=\rho_{{\rm
    PHF}}^{-}(r)$ (Fig.~\ref{den4}) agree very well with
ED~\cite{mikhailov} for large $r$. Generally, deviations occur near
$r\approx 0$.

\begin{figure}[htbp]
\includegraphics[width=8cm,keepaspectratio]{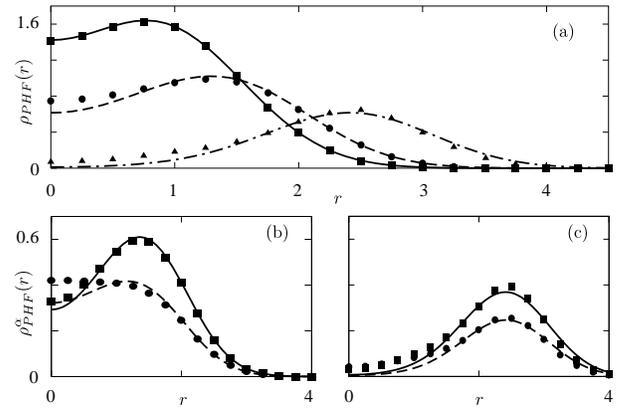}
\caption{\label{den5} Densities for  $N=5$ 
  $L=1$, $S=1/2$, and $S_{z}=1/2$. (a) Total density $\rho_{{\rm
      PHF}}(r)=\rho_{{\rm PHF}}^{+}(r)+\rho_{{\rm PHF}}^{-}(r)$ (units
  $\pi^{-1}\ell_{0}^{-2}$) for $\lambda=0.5$ (solid), $\lambda=2$
  (dashed), $\lambda=10$ (dashed-dotted). Squares, circles and
  triangles: data from CI.~\cite{rontani5} (b) Spin resolved densities
  $\rho_{{\rm PHF}}^{\alpha}(r)$ (solid: $\alpha=+$, dashed:
  $\alpha=-$, units $\pi^{-1}\ell_{0}^{-2}$) for
  $\lambda=2$. Squares, circles: data from CI.~\cite{rontani5} (c)
  Same as (b) but $\lambda=10$.}
\end{figure}

Figure~\ref{den5}(a) shows the total electron density for $N=5$,
$L,S=1,1/2$, for $\lambda=0.5,2,10$. 
For weak interaction, $\lambda=0.5$ (solid line), we
find good agreement with CI~\cite{rontani5} (squares). For $\lambda=2$
(dashed) small deviations near $r=0$ are found.
Figure~\ref{den5}(b) indicates that the spin-down density is
responsible for the small deviation from the exact result around $r=0$
for $\lambda=2$.
%%%%%%%%%%%%%%%%%%%%%%%%%%%%%%%%%%%%%%%%%%%%%%%%%%%%%%%%%%%%%%%%%%%%%%%%%%%%%%%
%
%
% SUBSECTION: Finite magnetic field
%
%
%%%%%%%%%%%%%%%%%%%%%%%%%%%%%%%%%%%%%%%%%%%%%%%%%%%%%%%%%%%%%%%%%%%%%%%%%%%%%%%
\subsection{Finite magnetic field}
\label{Bfinite}
In this section, we show results for $N=5,6,7$ in the presence of a
magnetic field, $B\leq 2.4\ {\rm T}$, corresponding to a dot filling
factor $\nu\gtrsim 2$ ($N\leq 4$ has been discussed in
Ref.~\onlinecite{degio}). We assume here a confinement $\omega_0=6$
meV (corresponding to $\lambda=1.45$) and $g^*=-0.44$. For $B>0$, due
to the Zeeman term, the PHF ground state always has $S_{z}=S$.
Therefore, we do not specify $S_{z}$ in the following.

\begin{figure}[htbp]
\includegraphics[width=8cm,keepaspectratio]{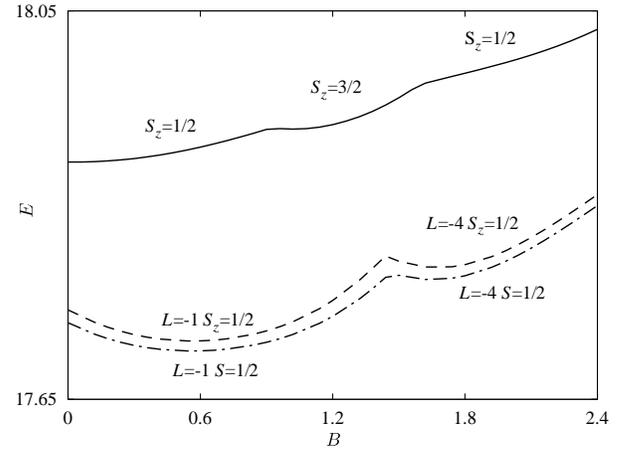}
\caption{\label{ground5} Ground state energy $E$ (units $\omega_0$) as
  a function of magnetic field $B$ (units T) for $N=5$. Solid: UHF;
  dashed: angular momentum projection; dashed-dotted: PHF.  Here, and
  in the following figures, $m^*=0.067\ m_e$, $\varepsilon_r=12.4$,
  $g^*=-0.44$, and $\omega_0= 6$ meV.}
\end{figure}

\begin{figure}[htbp]
\includegraphics[width=8cm,keepaspectratio]{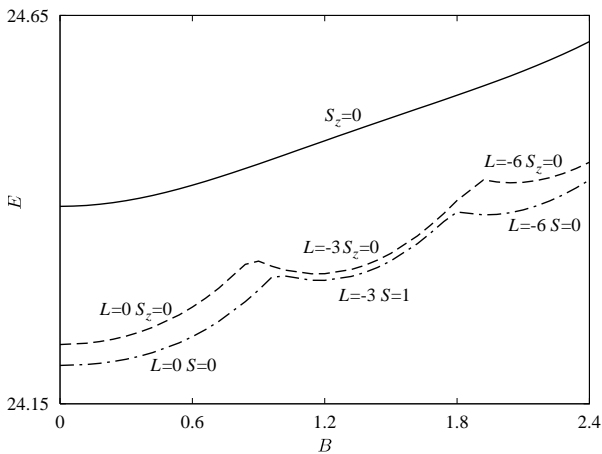}
\caption{\label{ground6} Ground state energy $E$ (units $\omega_0$) as a
  function of magnetic field $B$ (units T) for $N=6$. Solid: UHF;
  dashed: angular momentum projection; dashed-dotted: PHF.}
\end{figure}

We start with $N=5$ (Fig.~\ref{ground5}) and $N=6$
(Fig.~\ref{ground6}). We show the UHF ground state energy $E_{{\rm
    UHF}}$ (solid line), the energy obtained from angular momentum
projection (dashed, Eq.~(\ref{energyL})), and the PHF energy
(dashed-dotted, Eq.~(\ref{ephf})). The highest energy gain is here due
to the angular momentum projection. Spin projection leads to a further
decrease of the ground state energy. Obviously, UHF and PHF results
behave completely differently with $B$.

For instance, for $N=5$ (Fig.~\ref{ground5}) the UHF ground state
shows crossovers $S_{z}=1/2\to3/2$ at $B\approx 0.9\ {\rm T}$ and
$S_{z}=3/2\to1/2$ at $B\approx 1.5\ {\rm T}$. In contrast, the PHF
energy has total spin $S=1/2=S_{z}$ in the entire magnetic field
region. The state with $S_{z}=3/2$, not compatible with the total spin
$S=1/2$, is certainly an artifact of UHF. The crossover $L,S=1,1/2
\to L,S=4,1/2$ with increasing magnetic field at $B=1.4\ {\rm T}$
agrees quantitatively with the earlier results obtained by
ED.~\cite{tavernier}
\begin{figure}[htbp]
\includegraphics[width=8cm,keepaspectratio]{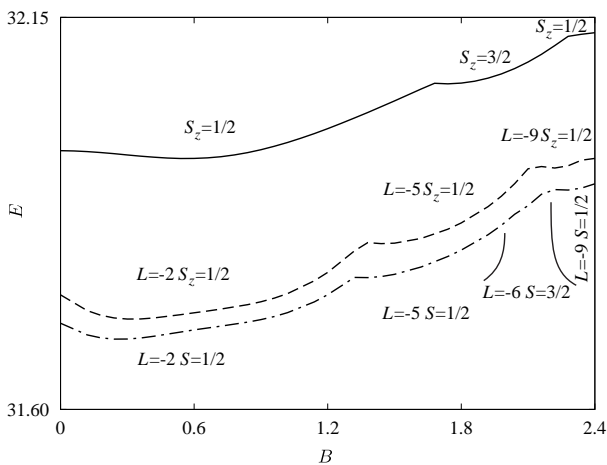}
\caption{\label{ground7} Ground state energy $E$ (units $\omega_0$) as a
  function of magnetic field $B$ (units T) for $N=7$. Solid: UHF;
  dashed: angular momentum projection; dashed-dotted: PHF. }
\end{figure}

For $N=6$ (Fig.~\ref{ground6}), UHF (solid line) displays no $S_{z}$
transitions. When rotational symmetry is restored, two crossovers,
({\em a}) $L=0\to -3$ and ({\em b}) $L=-3\to -6$ appear.  Performing
the spin projection, singlet states corresponding to $L=0$ and $L=-6$
are found, and $S=1$ for $L=-3$ is obtained. Singlets have the largest
energy gain, leading to a shift of the features found with angular
momentum projection (Fig.~\ref{ground6}). Also here, the PHF quantum
numbers agree with the earlier results obtained by
ED,~\cite{tavernier} including the magnitudes of the crossover fields
at $B\approx 1\ {\rm T}$ and $B \approx 1.8\ {\rm T}$ respectively.

The singlet-triplet crossover occurring for $N=6$ at $B\approx1.8\ 
{\rm T}$, corresponds to a filling factor $\nu\approx 2$, and is a
peculiar feature which is confirmed by several experimental and
theoretical studies.~\cite{hawrylak3,hawrylak4,hawrylak1,tarucha2}
Also for $N=8$ preliminary data indicate such a crossover near
$\nu=2$. These crossovers are completely absent in UHF
(Fig.~\ref{ground6}).

Most interesting is $N=7$ (Fig.~\ref{ground7}): near $B\approx2.2\ 
{\rm T}$ the ground state has $S=3/2$. This can only be obtained
including the spin projection and leads eventually to a spin blockade
in the transport (see below).

\begin{figure}[htbp]
\includegraphics[width=8cm,keepaspectratio]{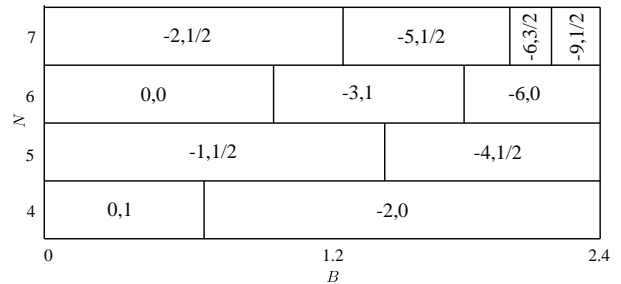}
\caption{\label{scheme} Scheme of the quantum numbers $L,S$ of the PHF
  ground state as a function of the magnetic field $B$ (units T) for
  $4\leq N\leq 7$.}
\end{figure}
In Fig.~\ref{scheme} we show the scheme of the ground state quantum
numbers for $4\leq N\leq7$, as obtained by PHF. They qualitatively
agree with previous calculations,~\cite{maksymB,tavernier} performed
for $N\leq 6$. In the region of $B$, where the ground state of $N=7$
has $S=3/2$ the state with $N=6$ is a singlet. Since $\Delta S>1/2$
between the two ground states a spin blockade in the
$6\leftrightarrows 7$ transition can be expected near the edge of
$\nu=2$ for $N=7$ electrons, for $B\approx2.3\ {\rm T}$.  We note in
passing that the lowest excited states for $N=7$, with $L=-5,S=1/2$ and
$L=-9,S=1/2$, are at most $\approx0.07$ meV ($\approx 0.8$ mK) higher
in energy.  Therefore, it may be difficult to experimentally observe
this blockade.

The chemical potential traces
$\mu_{N}(B)=E_{{\rm PHF}}(N,B)-E_{{\rm PHF}}(N-1,B)$ obtained by PHF
when varying $B$ are experimentally accessible via Coulomb blockade.
\begin{figure}[htbp]
\includegraphics[width=8cm,keepaspectratio]{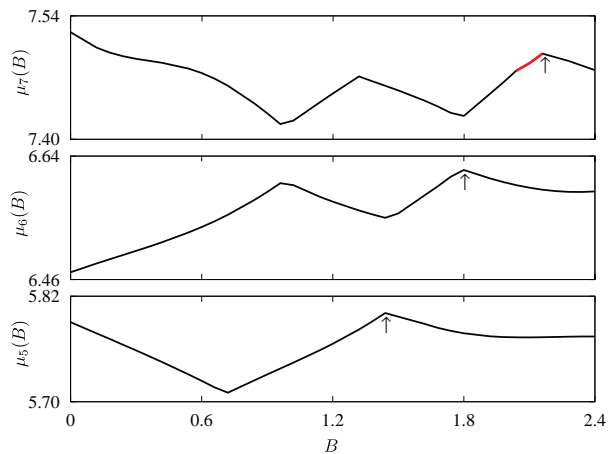}
\caption{\label{chemic} Chemical potentials
  $\mu_5(B)$, $\mu_6(B)$, $\mu_7(B)$ (units $\omega_0$) as a function
  of $B$ (unit T). Arrows: edge of filling factor $\nu=2$ for $N=5$ (bottom panel), 
  $N=6$ (center), $N=7$ (top).
   Red line: region of intrinsic spin blockade (see text).}
\end{figure}
Figure~\ref{chemic} shows $\mu_{5}(B)$, $\mu_{6}(B)$ and
$\mu_{7}(B)$. Arrows indicate the onset of $\nu=2$ for the
configuration with  $N=5$ (bottom panel), $N=6$ (center), $N=7$ (top).
The chemical potentials exhibit
features related to the above discussed crossovers between ground
states. At the onset of $\nu=2$, the chemical potentials exhibit a
cusp. For even $N$, this corresponds to the above mentioned
singlet-triplet transition.~\cite{hawrylak1} Generally, the chemical
potentials show kinks when quantum numbers of the ground states change
(Figs.~\ref{scheme} and~\ref{chemic}).
%%%%%%%%%%%%%%%%%%%%%%%%%%%%%%%%%%%%%%%%%%%%%%%%%%%%%%%%%%%%%%%%%%%%%%%%%%%%%%%
%
%
% SECTION: Conclusions
%
%
%%%%%%%%%%%%%%%%%%%%%%%%%%%%%%%%%%%%%%%%%%%%%%%%%%%%%%%%%%%%%%%%%%%%%%%%%%%%%%%
\section{Conclusion}
\label{conclusions}

We have described a systematic procedure to overcome some of the
limitations of UHF approach. Using angular momentum {\em and} total
spin projections, we have introduced correlations that provide
lower estimates for the ground state energies, besides determining the
spin and the angular momentum. Several sources of errors have been discussed.
In particular, a degeneracy error has been found to be useful for
deciding whether or not the estimate for the ground state is
plausible.

The procedure yields results consistent with earlier findings for
interaction strengths $\lambda\lesssim 2$ which corresponds to
experimentally relevant confinement energies $\omega_0\gtrsim 3$ meV for
$\varepsilon_{r}=12.4$.~\cite{kouwenhoven,tarucha}

For $B=0$ and $\lambda\leq2$, we have confirmed Hund's first rule for
the dot total spin, except for $N= 10$. In this case, the ground state
is ambiguous, since the energy gap between ground and first excited
state is smaller than the degeneracy error, consistent with other
results.  For stronger interaction, $\lambda>4$, deviations from
Hund's rules are obtained, accompanied by the well--known exchange
induced tendency of HF--based methods to favor ground states with
higher spins and zero angular momenta.

We have shown that PHF predicts correctly the features of the ground
state energy as a function of $B$. We have found a spin blockade in
the transport between $N=6$ and $N=7$, occurring at a filling factor
$\nu \approx 2$.

Given the slower increase in computational effort with particle number
described in Sec.~\ref{project} (Tab.~\ref{tab0}), as compared to
other methods, we hope by parallelization of our code to obtain in the
future results for higher number of particles ($N\geq 20$),
varying $B$, for interaction strengths relevant to quantum dot
experiments, $\lambda\leq 2$.

That the densities are correctly reproduced suggests that tunneling
rates between the quantum dot and attached leads needed for electron
transport can be reasonably well estimated when using PHF wave
functions. This might be useful for providing quantitative results for
predicting the heights of the Coulomb blockade peaks as a function of
$B$.\cite{roggekondo,hawrylak3,RHCSK2006}

\noindent
This work has been supported by the Italian MIUR via PRIN05, by the
European Union via MRTN-CT-2003-504574 contract and by the SFB 508
``Quantenmaterialien'' of the Universit\"at Hamburg.
\begin{appendix}
%%%%%%%%%%%%%%%%%%%%%%%%%%%%%%%%%%%%%%%%%%%%%%%%%%%%%%%%%%%%%%%%%%%%%%%%%%%%%%%
%
%
% APPENDIX: Some details of the implementation
%
%
%%%%%%%%%%%%%%%%%%%%%%%%%%%%%%%%%%%%%%%%%%%%%%%%%%%%%%%%%%%%%%%%%%%%%%%%%%%%%%%
\section{Some  details of  the implementation}
\label{details}
We provide some technical details about the implementation of the
projection technique outlined in Sec.~\ref{project}. In order to
obtain~(\ref{energyp}), we have to evaluate the overlaps
\begin{eqnarray}
&&\langle\Psi^{S_z}|\Psi^{S_{z}}(L,S)\rangle=\frac{1}{2\pi}
\int_{0}^{2\pi}{\rm d}\gamma\ e^{-iL\gamma}\cdot\nonumber\\
&&\sum_{q=0}^{N_{<}}\sum_{i=1}^{n_q}C_{q}(S,S_{z},N)\langle T_0|
T_{q}^{(i)}(\gamma)\rangle\, ,\label{D}
\end{eqnarray} and the Hamiltonian matrix elements
\begin{eqnarray}
&&\langle\Psi^{S_z}|H_{0}+V|\Psi^{S_{z}}(L,S)\rangle=
\!\frac{1}{2\pi}\int_{0}^{2\pi}{\rm d}\gamma\ e^{-iL\gamma}\cdot\nonumber\\
&&\sum_{q=0}^{N_{<}}\sum_{i=1}^{n_q}C_{q}(S,S_{z},N)\langle T_0|
H_0+V|T_{q}^{(i)}(\gamma)\rangle\, ,\label{Nh}
\end{eqnarray}
with
\begin{equation}
  H_{0}+V=\sum_{i=1}^{N}\left[h_{0}({\bf r}_{i},s_{zi})+\frac{1}{2}
\sum_{\substack{j=1\\j\neq i}}^{N}\!v({\bf r}_{i}-{\bf r}_{j})\!\right]\, .
\end{equation}
In the following, we avoid explicit reference to electron
coordinates unless when strictly necessary.

Previously, analytic expressions for~(\ref{energyp}) have been
reported~\cite{landman2} for $N=2$. For larger $N$, however,
the number and the complexity of the above matrix elements increases
dramatically. This eventually requires numerical treatment.

For evaluating
$|T_q\rangle=\sum_{i=1}^{n_{q}}|T_q^{(i)}\rangle$, we need to generate
all the $n_{q}$ swaps of $q$ opposite spin pairs in
$|T_0\rangle$. The latter correspond to a special class of
permutations, acting on the $k$-th component of the generalized
vector
\begin{equation}
{\bm \sigma}=\left(\chi^{+},\ldots,\chi^{+},\chi^{-},\ldots,\chi^{-}\right)\label{vecs}
\end{equation}
with the correspondence $k\to\pi_{k}^{(q,i)}$. One then has
\begin{equation}
  |T_q\rangle=\frac{1}{\sqrt{N!}}
\sum_{i=1}^{n_{q}}{\rm det}\{w_1\sigma_{\pi^{(q,i)}_1},\dots,w_N\sigma_{\pi^{(q,i)}_{N}}\}\, ,
\end{equation}
with
\begin{equation}
{\bm w}=\left(u_{1}^{+},\ldots,u_{N_{+}}^{+},u_{1}^{-},
\ldots,u_{N_{-}}^{-}\right)\, .\label{vecu}\\
\end{equation}
All the permutations are pre-tabulated at the beginning of the
calculation. Further calculations are performed by means of well known
theorems~\cite{loewdin2} for many body wave functions. The overlap
term is
\begin{equation}
\langle T_0| T_{q}^{(i)}(\gamma)\rangle={\rm det}\{d^{(q,i)}(\gamma)\}
\end{equation}
where $d^{(q,i)}$ is the overlap matrix
\begin{equation}
d^{(q,i)}_{kp}(\gamma)=\langle w_k|w_p(\gamma)\rangle \langle \sigma_k|\sigma_{\pi^{(q,i)}_p}\rangle\, .
\end{equation}
Here, $1\leq k,p\leq N$ and $w_{p}(\gamma)$ is a shorthand notation
for the rotated spatial part $w_{p}(r,\vartheta+\gamma)$. The term
$\langle\sigma_k|\sigma_{\pi^{(q,i)}_p}\rangle$ reduces to a Kronecker
delta. For the evaluation of $\langle w_k|w_p(\gamma)\rangle$ we use
the FD basis, which is particularly convenient to describe
rotations. The spatial parts transform as
\begin{equation}
  w_{p}(r,\theta+\gamma)=\sum_{\mu=1}^{K}
C_{\mu i}^{\alpha_{p}}e^{il_{\mu}\gamma}\phi_{\mu}(r,\theta)
\end{equation}
with $l_{\mu}$ the angular momentum of the $\mu$--th FD state,
$\alpha_{p}=+$ for $p\leq N_{+}$ and $\alpha_{p}=-$ for $N_{+}+1\leq p\leq
N$. Therefore,
\begin{equation}
\label{overlapping}
\langle w_k|w_p(\gamma)\rangle=
\sum_{\mu=1}^{K}(C_{\mu k}^{\alpha_k})^{*}C_{\mu p}^{\alpha_p}e^{il_{\mu}\gamma}
\end{equation}
which can easily be evaluated and stored.
  
The single particle term in the Hamiltonian is
\begin{equation}
\langle T_0|H_0|T_{q}^{(i)}(\gamma)\rangle=
\sum_{k,p=1}^{N}h_{0,kp}^{(q,i)}(\gamma)D^{(q,i)}_{k|p}(\gamma)
\end{equation}
where $D^{(q,i)}_{k|p}(\gamma)$ is the $k,p$ first order
cofactor of $d^{(q,i)}$ and
\begin{equation}
  h_{0,kp}^{(q,i)}(\gamma)=\langle  w_k ,\sigma_k | h_0|
w_p(\gamma) , \sigma_{\pi_{p}^{(q,i)}}\rangle\, .
\end{equation} 
In the interaction part
\begin{eqnarray}
\langle T_0|V|T_{q}^{(i)}(\gamma)\rangle&=&\frac{1}{2}
\sum_{k_1,p_1=1}^{N}\sum_{k_2,p_2=1}^{N}v^{(q,i)}_{k_1k_2p_1p_2}
(\gamma)\cdot\nonumber\\&\cdot&D^{(q,i)}_{k_1k_2|p_1p_2}(\gamma)\,, \label{tbt}
\end{eqnarray}
 $D^{(q,i)}_{k_1k_2|p_1p_2}(\gamma)$ represents the second order
cofactor of the matrix $d^{(q,i)}(\gamma)$, and
\begin{eqnarray*}
&&v^{(q,i)}_{k_1k_2p_1p_2}(\gamma)=\\
&&\langle w_{k_1}w_{k_2}|v|w_{p_1}(\gamma)w_{p_2}
(\gamma)\rangle\langle\sigma_{k_1}|\sigma_{\pi^{(q,i)}_{p_1}}\rangle
\langle\sigma_{k_2}|\sigma_{\pi^{(q,i)}_{p_2}}\rangle\,.
\end{eqnarray*}
Terms $H_{0,kp}^{(q,i)}(\gamma)$ and
$V^{(q,i)}_{k_1k_2p_1p_2}(\gamma)$ can be straightforwardly evaluated
as for~(\ref{overlapping}).

\end{appendix}

\end{document}